# Principles of perceptual grouping: implications for image-guided surgery

Birgitta Dresp-Langley

ICube UMR 7357 CNRS, University of Strasbourg (France)

E-mail: birgitta.dresp@icube.unistra.fr

## Abstract

*Gestalt* theory has provided perceptual science with a conceptual framework which has inspired researchers ever since, taking the field of perceptual organization into the 21st century. This opinion article discusses the importance of rules of perceptual organization for the testing and design of visual interface technology. It is argued that major *Gestalt* principles, such as the *law of good continuation* or the principle of *Prägnanz* (suggested translation: *salience*), taken as examples here, are important to our understanding of visual image processing by a human observer. Perceptual integration of contrast information across collinear space, and the organization of objects in the 2D image plane into figure and ground are of a particular importance here. Visual interfaces for image-guided surgery illustrate the criticality of these two types of perceptual processes for reliable decision making and action. It is concluded that Gestalt theory continues to generate powerful concepts and insights for perceptual science placed within the context of major technological challenges of today.

The laws and principles which predict how perceptual qualities can be extracted from the most elementary visual signals were discovered by *Gestalt* psychologists (e.g. Metzger, 1930, and Wertheimer, 1923, translated and re-edited by Spillmann in 2009 and 2012, respectively). Their seminal work has inspired visual science ever since, and has led to exciting discoveries which have confirmed the *Gestalt* idea that the human brain would have an astonishing capacity for selecting and combining critical visual signals to generate output representations for decision making and action. This capacity of selection and integration enables the perception of form and space, and the correct estimation of relative positions, trajectories, and distances of objects represented in planar images. The *Gestalt* laws and principles relate to the ability of human observers to assess 1) which parts of an image belong together to form a unified visual object or shape, and 2) which parts should be nearer and which further away from the observer if the represented objects were seen in the real world. Two of these, the principle of *Prägnanz* (suggested translation: salience) and the *law of good continuation* are considered here.

The *Gestalt* principle of *Prägnanz* governs specific perceptual processes that generate cues to shape and relative distance (figure-ground) in the visual field. These processes use local signals of contrast and orientation to fill in specific regions of an image and thereby enable the perception of surfaces. The associated perceptual sensations of local contrast enhancement make visual objects in the image appear to stand in front of other objects represented in the same plane. Such sensations are often deemed "illusory" because they have no physical origin, i.e. there is no objective difference in local luminance that would explain the resulting percepts (e.g. Heinemann, 1955; Hamada, 1985; O'Shea, Blackburn, and Ono, 1994; DeWeert & Spillmann, 1995; Grossberg, 1997; Dresp & Fischer, 2001; Dresp, Durand, and Grossberg, 2002; Guibal & Dresp, 2004; Devinck, Spillmann and Werner, 2006; Pinna & Reeves, 2006; Dresp-Langley & Durup, 2009; Dresp-Langley & Reeves, 2013, 2014). An essential aspect of this process of figure-ground segregation is the perceptual assignment of border ownership (see the review by von der Heydt on this topic). The *Gestalt* theorist Rubin (1921) was among the first to point out that a figure has distinct perceptual qualities that make it stand out against the rest of the visual field, which thereby acquires the perceptual quality of ground (or background). A figure occludes the ground and, therefore, owns the borders which separate it from the latter. Zhou, Friedman, and von der Heydt (2000) found neurons predominantly in V2 (but also V1) of the monkey that respond selectively to the location of borders in the visual field. Selective visual attention to the figure strengthens the neuronal responses to its borders (Qiu, Sugihar, and von der Heydt, 2007).

The *Gestalt* psychologists also correctly presumed that, to recover a representation of a whole from parts, the brain must achieve the perceptual integration of visual information across collinear space (e.g. Wertheimer, 1913; Metzger, 1930). The visual integration of contrast information across collinear image space plays a crucial role in form vision under conditions of stimulus uncertainty and configurative ambiguity (e.g. Dresp, 1997; Grossberg, 1997). It is governed by the so-called *law of good continuation*, and reflected by interactive effects between co-axial stimuli in the visual field. Specific response activities of visual cortical neurons are triggered by these co-axial interactions (cf. the first observations by Nelson & Frost, 1978 and von der Heydt, Peterhans, and Baumgartner, 1984 in monkey visual cortex), revealing the functional properties of brain mechanisms designed to complete physically discontinuous contrast input across collinear visual space. Collinear spatial integration is crucial for the detection of alignment, virtual trajectories, and shape borders in a world where most objects are seen incompletely. It enables a human observer to assess the continuity of image fragments under conditions of diminished visibility and heightened stimulus ambiguity. Experimental data on collinear visual integration have shown that the perceptual recovery of global representations of collinear space involves many levels of visual processing, not a single one, from the visual detection of local image detail to the perception of global association fields (e.g. Dresp, 1993; Field, Hayes, and Hess, 1993; Polat & Sagi, 1993, 1994; Kapadia et al., 1995; Polat & Norcia, 1996; Wehrhahn & Dresp, 1998; Yu & Levi, 1997, 2000; Chen, Kasamatsu, Polat, and Norcia, 2001; Chen & Tyler, 2001; Tzvetanov & Dresp, 2002; Dresp & Langley, 2005; Chen & Tyler, 2008; Huang, Chen, and Tyler, 2012; Spillmann, Dresp-Langley, and Tseng, 2015).

The laws of perceptual organization formulated by Gestalt theory have not lost any of their significance. They turn out to be as relevant as ever in the context of visual interface technology for image-guided surgery, for example. Image-guided surgery aims to use images taken before and/or during the procedure to help the surgeon navigate. The goal is to augment the surgeon's capacity for decision making and action during the procedure (see Perrin et al., 2009, for review). In augmented reality, the guidance is provided directly on the surgeon's view of the patient by mixing real and virtual images (Figure 1). This includes the visual tracking of devices relative to the patient, registration and alignment of the preoperative model, and the suitable rendering and visualization of the preoperative data. Visualization in this context means translating image data into a graphic representation that is understandable by the user (the surgeon), as it conveys important information for assessing structure and function, and for making (the right!) decisions during an intervention. The field has evolved dramatically in recent years, yet, the most critical problem for image-guided surgery is still the one of task-centred user interface design. During a surgical intervention, the timing of the generation of image data is absolutely critical, and to facilitate navigation through large cavities with multiple potential obstacles, such as within the abdomen, complex displays have been designed to provide navigational aids. They combine surface renderings of anatomy (Figure 1, left) from preoperative imaging with intra-operative visualization techniques. A common strategy here is representing volumetric data as 2D surfaces with varying opacity. The efficiency of renderings for facilitating decisions of the human user can be evaluated in terms of the perceptual salience of critical surfaces (principle of *Prägnanz*) that represent regions of interest to the surgeon.

Moreover, intra-operative imaging often provides further diagnostic information and permits assessing risks as well as perspectives of repair. In this context, image-guided instrument tracking is a major challenge for current research and development in this field (West et al., 2004; Huang et al., 2007). A critical problem for the surgeon is detecting and keeping track of the relative positions of the surgical tools he/she is using during the intervention (Figure 1, right). Visual tracking of the tooltip trajectories is also a precious aid for evaluating skill evolution in trainee surgeons, the positional accuracy of the tooltips being critical during an intervention (e.g. Jiang et al., 2014). The development and testing of new visual aids to facilitate the detection of alignment, relative position and trajectories (perceptual law of *good continuation*) is urgently needed here. Ultimately, technology where the surgical tool itself will become itself a visual navigation aid in image-guided surgery is to be developed in the near future and psychophysical testing should have a major impact on these developments.

**Figure caption**

<u>Figure 1</u>
In image-guided surgery, visual guidance is provided directly on the surgeon's view of the patient's anatomy by mixing real and virtual images. The efficiency of such renderings (**left**) for facilitating surgical decision and action is critically determined by the salience of the rendered surfaces (principle of *Prägnanz*) that represent regions of interest to the surgeon. Visual tracking of the tooltip trajectories is important for evaluating skill evolution, the positional accuracy of the tooltips being critical. Technology facilitating the positional accuracy of tool-tip movements (**right**) by generating visual data for relative position, alignment, and trajectory anticipation (perceptual law of *good continuation*) is needed.

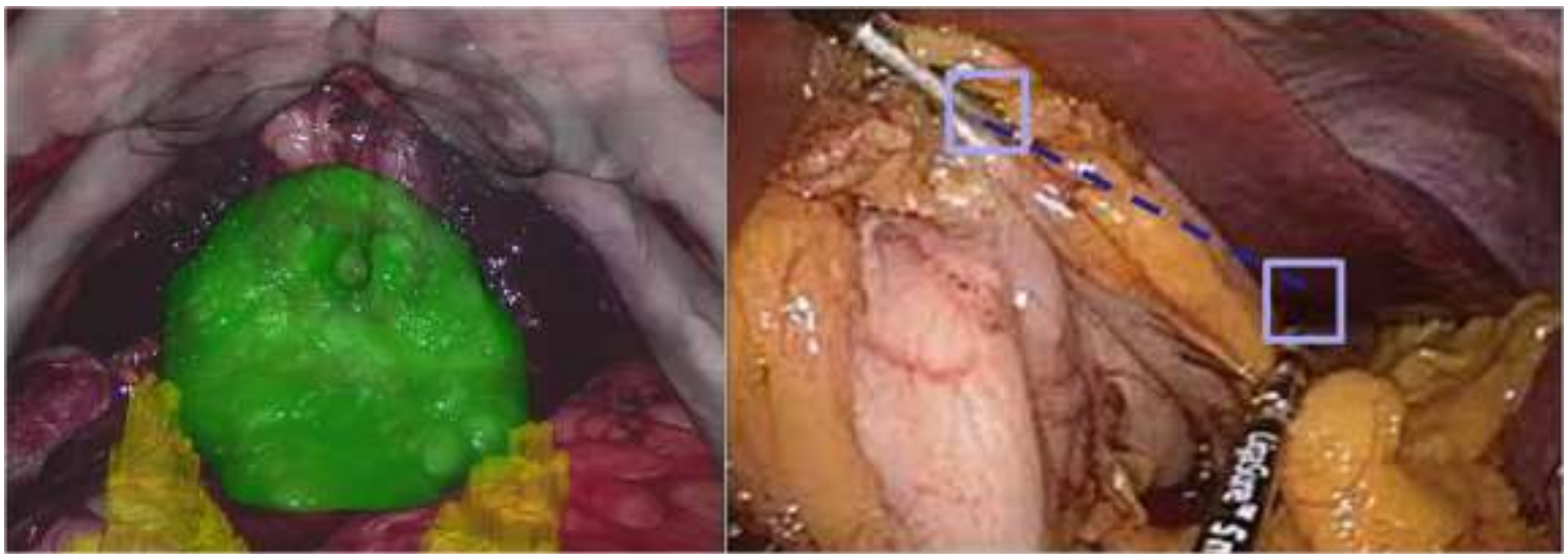

Figure 1